# Disentangling and quantifying market participant volatility contributions

Marcello Rambaldi*†, Emmanuel Bacry†§, and Jean-François Muzy‡

†CEREMADE, Université Paris-Dauphine, CNRS, UMR 7534, 75775 Paris, France
‡SPE, Université de Corse, CNRS, UMR 6134, Campus Grimaldi, 20250 Corte, France
§CMAP, Ecole Polytechnique, CNRS, UMR 7641, 91128 Palaiseau, France

Thanks to the access to labeled orders on the Cac40 index future provided by Euronext, we are able to quantify market participants contributions to the volatility in the diffusive limit. To achieve this result we leverage the branching properties of Hawkes point processes. We find that fast intermediaries (e.g., market maker type agents) have a smaller footprint on the volatility than slower, directional agents. The branching structure of Hawkes processes allows us to examine also the degree of endogeneity of each agent behavior. We find that high-frequency traders are more endogenously driven than other types of agents.

## 1. Introduction

During the past two decades, with the availability of an increasing amount of high-frequency market data, significant progresses have been achieved in the empirical characterization of orderbook dynamics and the main features related to price formation (see e.g. Bouchaud et al. (2018) for a recent review): the long-range correlated nature of supply and demand, the concave law for the impact of orders and meta-orders, the latent character of the available liquidity or the strong endogenous (reflexive) nature of price variations are examples of properties that have been empirically highlighted in many exchanges. From a theoretical point of view, among the numerous stochastic models proposed to reproduce the order book features from econophysics or queuing theory (Smith et al. (2003), Cont et al. (2010), Abergel and Jedidi (2013), Huang et al. (2014)), Hawkes processes have proven to provide a very appealing alternative since they constitute a tractable class of multivariate point processes that allow one to account for the self and cross interactions of various kind of events (see Bacry et al. (2015) and references therein). Hawkes processes notably provide a suitable framework to establish granger-causal relationships between the occurrence of order-book events (Hongteng et al. (2016), Achab et al. (2017, 2018)) and to disentangle exogenous versus endogenous activity (Filimonov and Sornette (2012), Hardiman et al. (2013), Bacry et al. (2016)).

In this paper, our goal is to enrich former multivariate Hawkes models of Level-I order book events (as in Bacry et al. (2016), Achab et al. (2018)) by accounting for the identity of market participants. This is made possible by exploiting a labeled database provided by Euronext containing all the orders sent to the CAC 40 index future orderbook during the period extending from March 2016 to February 2017. Each order is labeled by the (anonymized) ID of the market member who sent this order. We associate, with each event type at level I of the order book (mid-price jumps, market/limit/cancel orders at bid/ask sides) initiated by a given agent, a component of a large dimensional Hawkes process (thus the overall dimension is the number of event types times the number of agents). By calibrating such a model on our dataset, we aim at characterizing the behavior of various market participants and their interactions. More specifically, we characterize the relative contribution of each agent action to the overall long term (e.g. daily) volatility with the aim of establishing an agent classification based on this dynamical feature. We will also try to quantify, as naturally allowed within the framework of Hawkes processes (Filimonov and Sornette (2012), Hardiman et al. (2013), Bacry et al. (2016)), for various market conditions and for different class of agents, in what proportion the activity is of reflexive nature (endogenous) as opposed to driven by some external information flow (exogenous).

*Corresponding author. Email: marcello.rambaldi@polytechnique.edu

The paper is organized as follows: In Section 2, we recall how the long range volatility can be directly determined within a multivariate Hawkes model that accounts for the mid-price variations. In Section 3, we define a "multi-agent" Hawkes model where each component is associated with a given type of level-I book order for given market participant. We discuss how such a high dimensional model can be calibrated provided one assumes that the reaction of some agent to an event initiated by another agent does not depend on the latter. In Section 4, we describe the Euronext future CAC40 labeled database and we provide some basic statistics that characterize the 16 most active market participants. Empirical results obtained from our model calibration are discussed in Section 5. We report some results about agent contribution to the daily volatility and also about agent reflexivity level. Moreover, the volatility contribution of each endogenous/exogenous agent activity is quantified in various volatility environment (highly volatile versus calm time periods). Concluding remarks and prospects for future research along the line drawn in this paper are provided in Section 6.

## 2. Measuring volatility contributions with Hawkes processes

### 2.1. *Introduction to Hawkes processes and Notations*

A multivariate Hawkes process of dimension $n$ is a $n$-dimensional counting processes $\mathbf{N}_t$ with a conditional intensity vector $\{\lambda_t^i\}_{1 \leq i \leq n}$ that is a linear function of past events. More precisely,

$$\lambda_t^i = \mu^i + \sum_{j=1}^{n} \int_{-\infty}^{t} \phi^{ij}(t-s) dN_s^j, \tag{1}$$

where $\mu^i$ stands for the baseline (exogenous) intensity of the $i$th component and $\phi^{ij}(t)$ is the Hawkes kernel function (positive and supported by $\mathbb{R}^+$) that quantifies the excitation rate of an event of the $j$th component on the arrival rate of events of the $i$th component after a time lag $t$.

Let $\boldsymbol{\Phi}(t)$ denote the matrix of kernels $\phi^{ij}(t)$ and set

$$\mathbf{R}(t) = \sum_{n \geq 0}^{\infty} \boldsymbol{\Phi}^{\star n}(t) \tag{2}$$

where $\boldsymbol{\Phi}^{\star k}(t)$ stands for $\boldsymbol{\Phi} \star \boldsymbol{\Phi} \star \ldots \star \boldsymbol{\Phi}(t)$ with $\boldsymbol{\Phi}$ repeated $k$ times (where we use the convention $\boldsymbol{\Phi}^{\star 0}(t) = \mathbf{Id}$) and $\star$ stands for a regular matrix multiplication in which each elementary multiplication is replaced by a convolution. Thus if $\mathbf{A}(t)$ and $\mathbf{B}(t)$ are square matrices whose elements are functions, the $i, j$ element of the matrix $\mathbf{A} \star \mathbf{B}(t)$ writes $\int \sum_k A^{ik}(s) B^{kj}(t-s) ds$.

If we denote $\boldsymbol{\Phi}$ and $\mathbf{R}$ the matrices obtained by considering the integrals of the components of respectively $\boldsymbol{\Phi}(t)$ and $\mathbf{R}(t)$ (i.e. $\Phi^{ij} = \int_0^{\infty} \phi^{ij}(t) dt$ and $R^{ij} = \int_0^{\infty} R^{ij}(t) dt$) then one can easily

prove that:

$$\mathbf{R} = \mathbf{Id} + \sum_{n \geq 1} \mathbf{\Phi}^n = (\mathbf{Id} - \mathbf{\Phi})^{-1} \,. \tag{3}$$

Let us remark that, as it follows from the cluster (or branching) representation of Hawkes processes, $\Phi^{ij}$ represents the mean total number of events of the $i$th component directly triggered by an event of the $j$th component. For this reason the matrix $\mathbf{\Phi}$ is also referred to in the literature as the branching ratio matrix. In the same way $R^{ij}$ represents the average number of events of the $i$th component triggered (directly or indirectly) by an exogenous event of the $j$th component (including the exogenous events themselves in the case $i = j$).

When one analyzes empirical data within the framework of Hawkes processes, the previous remarks allow one to quantify causal relationships between events in the sense of Granger, i.e., within a well defined mathematical model. In this respect, the coefficients of the matrices $\mathbf{\Phi}$ or $\mathbf{R}$ can be read as (Granger-)causality relationships between various types of events and used as a tool to disentangle the complexity of the observed flow of events occurring in some experimental situations (Achab et al. (2017)).

Using these notations, first and second order cumulants densities associated with the components $dN_t^i$, can be easily formulated (Jovanović et al. (2015), Bacry and Muzy (2016)). The unconditional intensity $\Lambda^i dt = E(dN_t^i)$ can be decomposed using the previously mentioned branching representation (i.e., each event has an exogenous event as its oldest ancestor):

$$\Lambda^i = \sum_k R^{ik} \mu^k \tag{4}$$

while the infinitesimal covariance matrix $C^{ij}(t)dtds = E(dN_s^i dN_{s+t}^j) - \Lambda^i \Lambda^j dtds$ can be linked with the matrix $R^{ij}(t)$ through the relation

$$C^{ij}(t) = \sum_m \Lambda^m \int R^{im}(u) R^{jm}(t+u) du \,. \tag{5}$$

### 2.2. *Measuring volatility contributions*

The so-called *microstructure noise* and the behavior of the *signature plot* were, for the first time, modeled using Hawkes processes in Bacry et al. (2013). In this work the authors considered upward and downward mid-price changes as events of a multivariate counting process. In this section, we show how, in such models, the value of the long term ("diffusive") volatility can be linked to the high-frequency price process.

Let $\mathbf{N}_t$ be a $n$-dimensional Hawkes process where some components $i \in \mathcal{M}$ explicitly account for mid-price move events. Thus, each of these components is characterized by the direction of the corresponding price-jump (upward or downward) and its size (one or several half-ticks). For $i \in \mathcal{M}$, let $\delta_i$ denote the signed amplitude of the mid-price jump associated with mid-price move events of type $i$, then the mid-price variation between times $t$ and $t + \tau$ reads

$$\Delta_\tau P(t) \equiv P(t+\tau) - P(t) = \sum_{i \in \mathcal{M}} \delta_i \int_t^{t+\tau} dN_s^i, \tag{6}$$

and the natural "no trend" condition $E(\Delta_\tau P) = 0$ gives

$$\sum_{i \in \mathcal{M}} \delta_i \Lambda^i = 0 \tag{7}$$

From (6), $\sigma_\tau^2 = E(\Delta_\tau P^2)$, the price squared volatility at time scale $\tau$, is obtained as:

$$\sigma_\tau^2 = \sum_{i,j \in \mathcal{M}} \delta_i \delta_j \int_0^\tau \int_0^\tau E(dN_s^i dN_{s'}^j) \tag{8}$$

which directly gives, using Eq. (7),

$$\sigma_\tau^2 = \sum_{i,j \in \mathcal{M}} \delta_i \delta_j \int_0^\tau \int_0^\tau C^{ij}(s-s') ds ds'. \tag{9}$$

By setting $x = s + s'$ and $y = s - s'$, one obtains (since the determinant of the Jacobian is $1/2$):

$$\sigma_\tau^2 = \frac{1}{2} \sum_{i,j \in \mathcal{M}} \delta_i \delta_j \iint_{(x,y) \in L} C^{ij}(y) dx dy \tag{10}$$

where $L$ is the image of the square $[0,\tau] \times [0,\tau]$ by the above change of variables, i.e., the domain $(-\tau \leq y \leq 0, -y \leq x \leq y + 2\tau) \cup (0 \leq y \leq \tau, y \leq x \leq 2\tau - x)$. After a little algebra, one can see that all integrals in the previous expression can be written as

$$\iint_{(x,y) \in L} C(y) dx dy = \tau \int_{-\tau}^\tau C(y) dy - \int_0^\tau y \left[C(y) + C(-y)\right] dy \tag{11}$$

It follows that, in the *large scale limit* $\tau \to \infty$, provided $C^{ij}(x)$ decreases sufficiently fast, one obtains from Eqs. (11) and (5):

$$\frac{\sigma_\tau^2}{\tau} \to \sum_{i,j \in \mathcal{M}} \delta_i \delta_j C^{ij} = \sum_{m=1}^n \Lambda^m \sum_{i,j \in \mathcal{M}} \delta_i \delta_j R^{im} R^{jm} = \sum_{m=1}^n \Lambda^m \left(\sum_{i \in \mathcal{M}} \delta_i R^{im}\right)^2 \tag{12}$$

This represents the expression of the price squared volatility at large scale. Let us remark that due to the identity term in $R^{ij}$ (see Eq. (2)) the previous expression cannot be zero. The final result is particularly simple, since the large scale squared volatility is just the weighted sum over each type of event (the weights being their intensity) of the square of some "mean" price jump they trigger (since $R^{im}$ is the mean total number of events of type $i$ triggered by the occurrence of an event of type $m$). In this respect the contribution of each component to the volatility is explicit. We note that within this "large scale" framework must be understood as large with respect to the support of the covariance functions.

As a simple example, one can consider the following toy model: we set $n = 2$ and $\mathcal{M} = \{1, 2\}$ (price up $\delta_1 = 1$ and price down $\delta_2 = -1$) and choose a fully symmetric $\mathbf{\Phi}$ matrix with $\Phi^{11} = \Phi^{22} = \varphi_S$ (the subscript $S$ stands for *self*) and $\Phi^{12} = \Phi^{21} = \varphi_C$ (the subscript $C$ stands for *cross*). In that case, the matrix $\mathbf{R}$ reads:

$$\mathbf{R} = \begin{bmatrix} \frac{1-\varphi_S}{(1-\varphi_S)^2-\varphi_C^2} & \frac{\varphi_C}{(1-\varphi_S)^2-\varphi_C^2} \\ \frac{\varphi_C}{(1-\varphi_S)^2-\varphi_C^2} & \frac{1-\varphi_S}{(1-\varphi_S)^2-\varphi_C^2} \end{bmatrix} \tag{13}$$

If both baseline intensities are $\mu^1 = \mu^2 = \mu$, we have $\Lambda^1 = \Lambda^2 = \Lambda = \frac{\mu}{1-\varphi_S-\varphi_C}$ and Eq. (12) gives:

$$\sigma^2 = \frac{2\Lambda}{(1-\varphi_S+\varphi_C)^2} = \frac{2\mu}{(1-\varphi_S-\varphi_C)(1-\varphi_S+\varphi_C)^2} \ . \tag{14}$$

Let us notice that in the case $\varphi_S = 0$, i.e., when one only considers cross exciting kernels, one recovers the expression originally obtained in (Bacry et al. (2013)). Let us also notice that $2\Lambda$ in Eq. (14) represents the long term squared volatility obtained when upward and downward price jumps are provided by two independent homogeneous Poisson processes of intensity $\Lambda$. Therefore, compared to a pure Poisson situation, the cross exciting term $\varphi_C$ decreases the volatility while the self-exciting term $\varphi_S$ increases it. For a general Hawkes process, both scenarios can thus occur.

## 3. A multi-agent model

In refs. (Bacry et al. (2016), Achab et al. (2018)), the authors introduced a Hawkes model to account for the various self/cross excitation/inhibition relationships that govern the dynamics of level-I order book events. As mentioned in the introduction, our aim is to refine this approach in order to account for the specific activity of each market participant. For this purpose, we consider a large dimensional Hawkes model that is able to account for all agent actions and interactions within an order book at level-I. In this way, we will be able to estimate the contributions of individual agents to the total asset volatility.

Let $\mathcal{A}$ be a set of $M$ agents ($|\mathcal{A}| = M$) and consider the set of order types $\mathcal{T} = \{P^+, P^-, L^a, L^b, C^a, C^b, T^a, T^b\}$ ($|\mathcal{T}| = 8$) where

- $P^+$ ($P^-$) are orders that immediately move upward (downward) the mid-price (whatever the size of the jump is)
- $T^a$ ($T^b$) are aggressive orders that are executed immediately at the best ask (bid) and do not move the mid price;
- $L^a$ ($L^b$) are new limit orders that arrive at the best ask (bid) (so they do not move the mid-price)
- $C^a$ ($C^b$) are cancellations of orders at the best ask (bid) that do not empty the queue (so again they do not move the mid-price)

In what follows, we will denote by $N_{i,\alpha}(t)$ the counting process that is associated with orders of type $\alpha \in \mathcal{T}$ placed by agent $i \in \mathcal{A}$. Moreover, we suppose that all the $8M$ components are described by a multivariate Hawkes process, i.e., the conditional intensity of $N_{i,\alpha}$ reads:

$$\lambda_t^{i,\alpha} = \mu^{i,\alpha} + \sum_{j\in\mathcal{A}}\sum_{\beta\in\mathcal{T}} \int_0^t \phi^{i,a;j,\beta}(t-s)dN_s^{j,\beta} \tag{15}$$

where $\mu^{i,\alpha}$ is the baseline intensity and the interaction kernel $\phi^{i,a;j,\beta}(s)$ represents the impact of an event of type $\alpha$ of agent $j$ on the occurrence likelihood of an event of type $\alpha$ of agent $i$ after a delay $s$. In order to estimate such a model, one can use a parametrized version of previous equation where each kernel is decomposed over a fixed dictionary of $L$ function $g_\ell(s)$:

$$\phi^{i,\alpha;j,\beta}(s) = \sum_{\ell=1}^{L} \theta_\ell^{i,\alpha;j,\beta} g_\ell(s) \ . \tag{16}$$

Within this framework we have $(8ML+1)8M$ parameters to estimate, a number that is large when there is a large number of interacting agents ($M \gg 1$). In order to estimate the model, one can use, as proposed by Hansen et al. (2015), a least square method that amounts to minimize the following contrast function:

$$\mathcal{C} = \sum_{i \in \mathcal{A}, a \in \mathcal{T}} \mathcal{C}^{i,\alpha}$$

where

$$\mathcal{C}^{i,\alpha} = T^{-1} \int_0^T \left(\lambda_s^{i,\alpha}\right)^2 ds - \frac{2}{T} \int_0^T \lambda_s^{i,\alpha} dN_s^{i,\alpha} \; .$$

In order to handle the problem complexity, Hansen et al. (2015) considered a lasso regularization that encodes a sparsity prior. In our problem, such a penalization is not natural since there is no particular reason to expect a sparse interaction matrix between agents. It is however natural to assume another prior: the way that some agent $i$ reacts to the action $\beta$ of another agent $j$ does not depend on $j$ unless $i = j$. This results from the fact that any agent perceives the activity of other agents only through their anonymous orders in the book. This hypothesis amounts to assume that:

$$\theta_\ell^{i,\alpha;j,\beta} = \begin{cases} \alpha_\ell^{i,\alpha;\beta} \text{ if } i = j \\ \beta_\ell^{i,\alpha;\beta} \text{ otherwise.} \end{cases} \tag{17}$$

From Eqs. (15) and (16), it results that $\lambda_t^{i,\alpha}$ can be rewritten as:

$$\lambda_t^{i,\alpha} = \mu^{i,\alpha} + \sum_{\ell,\beta} \alpha_\ell^{i,\alpha;\beta} N_\ell^{i,\beta}(t) + \sum_{\ell,\beta} \beta_\ell^{i,\alpha;\beta} M_\ell^{i,\beta}(t) \tag{18}$$

where

$$N_\ell^{i,\beta}(t) = \int_0^t g_\ell(t-s) dN_s^{i,\beta}$$

represents the activity of agent $i$ with orders of type $\beta$ filtered through the element $g_\ell$ of the dictionary while

$$M_\ell^{i,\beta}(t) = \sum_{j \neq i} \int_0^t g_\ell(t-s) dN_s^{j,\beta}$$

represents the $g_\ell$-filtered flow of orders of type $\beta$ due to the activity of other agents, i.e., associated with the "market" activity.

Let us remark that, since

$$\underset{\{\mu^{i,\alpha}, \theta_\ell^{i,\alpha;j,\beta}\}_{i,j,\alpha,\beta,\ell}}{\arg\min} \mathcal{C} = \bigcup_i \underset{\{\mu^{i,\alpha}, \theta_\ell^{i,\alpha;j,\beta}\}_{j,\alpha,\beta,\ell}}{\arg\min} \sum_\alpha \mathcal{C}^{i,\alpha} \; ,$$

Equation (18) means that the model parameters $\{\mu^{i,\alpha}, \alpha_\ell^{i,\alpha;\beta}, \beta_\ell^{i,\alpha;\beta}\}_{\alpha,\beta \in \mathcal{T}, \ell \in \{1...L\}}$ of a given agent $i$ can be estimated as a $2 \times 8 = 16$ dimensional model where one considers all events types associated with the agent on one hand and the "market" (i.e. the sum over all other agents without distinction) on the other hand. The estimation of the original model of dimension $8M$ is thus decomposed in a set of $M$ estimations of models of dimension $2 \times 8 = 16$.

## 4. Data and summary statistics

In this work, we use data on the CAC40 index future from Euronext. The raw data consist of all orders and trades submitted[1] on Euronext trading platform in the period from March 1st 2016 to February 28th 2017 for all contract expiries[2]. A member is an institution that has direct access to the market. Each member can send orders to the market through one or several connections. Our database allows us to reconstruct the full order book and, for each order, features a numeric ID indicating the member who submitted the order and a second ID identifying the particular connection used. Let us mention that both IDs are anonymized, so we do not have access to the actual identity of the members. In our dataset we have record of 111 unique members. The number of connections used by each member varies greatly, ranging from more than one hundred to just one.

As noted above, all contract maturities are present in our data as well as calendar spreads. For a given day the front maturity is considered to be the contract with the closest expiry date, except if the expiry date is on the same day[1]. Since the front maturity is by far the most represented in terms of orders and trades (more than 99% of orders and trades except during the 3 days before the expiry date), in the following we will consider only orders and trades on the front maturity.

In this study we focus on members active throughout the continuous trading phase that overlaps with the one of European equity markets. The reason for this choice is that most of the market activity concentrates during this time slot. Moreover, since we aim at fitting point processes models on agents data, we restrict our analysis to members that have at least 1000 orders at the best quotes during this period. Finally, in order to have a sufficient number of points for each member, we consider the member IDs that respect the aforementioned conditions for at least 30 trading days. This leaves us with 16 member IDs. For each member, we compute several features for each day in our sample. The chosen metrics allow us to roughly categorize the behavior of the selected agents. Note that these features are similar to those employed in the recent literature on agent and high-frequency traders characterization (as e.g. in Kirilenko et al. (2017), Menkveld (2013), ESMA Economic Report No. 1 (2014), Hagströmer and Nordén (2013), Brogaard et al. (2014), Bellia (2017), Biais and Foucault (2014), Megarbane et al. (2017)). The detailed description for each feature we considered is provided in Table 1. For each agent, we computed the average value of each feature over all days during which the agent was active according to our criteria, i.e. he passes at least 1000 orders at the best quotes. All these average values are displayed in Table 2. For the sake of clarity, members are sorted from left to right according to the first feature (i.e., their average position change at the end of the day).

Let us note that the members on the left side of the table are those corresponding to the profile of high-speed intermediaries: they end the day almost perfectly flat, are fast at canceling/modifying their outstanding orders and are highly present at the best quotes. In the following, we will emphasize on these four features to distinguish market participants, namely the EOD position, the presence at both best quotes, the average order lifetime and the fraction of traded volume that is aggressive. Besides considering the agents individually, we will consider conditional averages over quantiles of these 4 features. This will allow us to smoothly go from "market-marker" type behavior (quantiles around zero) to "directional traders" type activity (quantiles with high values).

---

[1] With the exception of orders refused by the market upon submission, e.g. Fill Or Kill orders that are not filled.

[2] For the CAC40 future there are 3 monthly, 3 quarterly (from March, June, September, December), and 8 half-yearly maturities from the June/December cycle

[1] Trading ceases at 16:00 CET on the third Friday of the delivery month. In the event that the third Friday is not a business day, the Last Trading Day shall normally be the last business day preceding the third Friday

July 18, 20188

| Feature name | Feature description |
| --- | --- |
| End of day (EOD) position | Absolute change in inventory at the end of the trading day, divided by the total volume traded by the agent. |
| Proprietary | Fraction of the orders that are market as proprietary by the agent. |
| Order lifetime | Median time between limit order insertion and cancellation/modification. |
| Inter-event time | Median time between two different orders by the same agent. |
| Limit-filled | Fraction of the submitted limit orders that are at least partially filled. |
| Canceled orders | Fraction of limit orders that are eventually canceled. |
| Aggressive volume | Ratio of the volume traded aggressively over the total traded volume by the agent. |
| Orders/Trades | Number of orders submitted for each trade. |
| Order size | Average order size (in contracts). |
| Time present at L1 | Fraction of time the agent was present with a limit at at least one of the best quotes. |
| Present at both sides | Given the agent was present at the best, fraction of time he was present at both sides simultaneously. |
| Active connections | Average number of connections used by the agent per day. |
| Daily volume fraction | Fraction of the total traded volume (total buy + total sell) in which the agent is involved. |

Table 1. Feature description. All features are computed per-member.

| | 240 | 140 | 478 | 127 | 636 | 398 | 503 | 274 | 566 | 59 | 584 | 364 | 597 | 455 | 244 | 669 |
|---|---|---|---|---|---|---|---|---|---|---|---|---|---|---|---|---|
| EOD Position / Volume (%) | 0.00 | 0.01 | 0.15 | 3.73 | 3.83 | 4.54 | 9.71 | 14.9 | 16.2 | 22.3 | 18.3 | 22.7 | 24.5 | 29.1 | 28.2 | 32.8 |
| % Proprietary | 100.0 | 100.0 | 100.0 | 100.0 | 100.0 | 100.0 | 0.22 | 68.1 | 97.8 | 100.0 | 1.19 | 100.0 | 2.10 | 98.7 | 0.00 | 0.37 |
| Order lifetime (s) | 0.51 | 0.61 | 0.20 | 3.57 | 0.99 | 0.33 | 1.33 | 3.19 | 42.0 | 4.14 | 7.87 | 5.17 | 4.32 | 3.04 | 6.31 | 11.1 |
| Inter-event time (s) | 0.01 | 0.00 | 0.02 | 0.01 | 0.00 | 0.06 | 0.63 | 0.07 | 0.63 | 0.01 | 1.64 | 0.01 | 1.65 | 0.12 | 2.45 | 2.33 |
| Limit filled (%) | 5.09 | 6.15 | 8.40 | 10.5 | 6.35 | 10.5 | 28.3 | 19.8 | 47.9 | 1.58 | 50.4 | 5.45 | 42.4 | 4.05 | 23.5 | 42.0 |
| Limit (%) | 51.1 | 50.0 | 48.9 | 44.3 | 36.3 | 37.7 | 49.3 | 47.5 | 53.6 | 40.7 | 31.0 | 51.0 | 54.1 | 50.0 | 48.1 | 53.4 |
| Canel (%) | 48.4 | 47.2 | 46.2 | 40.0 | 33.7 | 33.9 | 36.2 | 37.9 | 27.9 | 40.1 | 14.5 | 48.3 | 31.1 | 48.0 | 36.6 | 30.2 |
| Replace (%) | 0.00 | 0.08 | 3.43 | 13.6 | 29.4 | 27.4 | 6.58 | 8.78 | 7.54 | 18.4 | 40.1 | 0.04 | 5.77 | 1.57 | 11.1 | 8.57 |
| Aggressive (%) | 0.51 | 2.69 | 1.42 | 2.08 | 0.60 | 1.01 | 7.97 | 5.76 | 10.9 | 0.80 | 14.4 | 0.62 | 9.08 | 0.43 | 4.25 | 7.78 |
| Aggressive volume (%) | 14.9 | 64.0 | 34.0 | 34.4 | 15.0 | 13.2 | 49.9 | 46.9 | 37.4 | 56.2 | 44.4 | 25.0 | 34.8 | 27.0 | 27.0 | 28.8 |
| Orders/Trades (%) | 3994.2 | 1085.8 | 1351.5 | 1128.0 | 5573.1 | 1036.1 | 238.5 | 524.7 | 190.3 | 5276.6 | 191.9 | 2609.4 | 206.5 | 3915.7 | 162.6 | 497.4 |
| Order size (contracts) | 1.02 | 1.38 | 2.33 | 1.65 | 1.15 | 4.41 | 2.45 | 1.64 | 1.70 | 1.88 | 3.66 | 2.42 | 2.70 | 4.08 | 3.75 | 2.38 |
| Time present at L1 (%) | 76.8 | 99.4 | 51.1 | 87.6 | 73.7 | 26.5 | 39.3 | 38.4 | 22.7 | 30.4 | 19.7 | 36.1 | 25.1 | 22.2 | 27.0 | 42.6 |
| Present at both sides (%) | 39.1 | 69.1 | 9.21 | 36.9 | 25.9 | 0.69 | 4.71 | 5.07 | 1.59 | 1.61 | 0.99 | 1.32 | 1.75 | 0.69 | 1.91 | 5.87 |
| Active connections | 19.9 | 98.2 | 16.2 | 32.2 | 19.6 | 2.16 | 18.9 | 19.8 | 9.32 | 5.47 | 17.7 | 10.5 | 4.26 | 13.9 | 2.55 | 3.69 |
| Daily volume fraction (%) | 2.22 | 31.3 | 4.68 | 6.30 | 1.28 | 3.59 | 6.05 | 5.63 | 2.04 | 4.76 | 3.85 | 2.00 | 1.88 | 2.13 | 2.65 | 2.73 |

Table 2. Average features of the selected 16 agents. The agents are sorted from left to right on the first feature, i.e., their average closing position (leftmost is flatter). All features are calculated on each day during European equity hours (08:00 - 16:30 London time) except for the position which is calculated on the whole trading day (07:00 - 21:00).

## 5. Empirical results

We consider the approach outlined in the previous sections to quantify and analyze the contribution to the diffusive volatility of the various market participants. For each day in the sample and for each operating agent, we estimate a model where we distinguish two actors, the agent and the rest of the market. For each actor, we separate components corresponding to orders that move the mid-price (one component for upward moves and one for downward moves) as well as components for limit, cancel and aggressive orders (2 components bid/ask for each) that do not move the mid-price, as outlined in Section 3. We thus have 8 components for each actor by distinguishing up/down price movements and orders that hit the bid or the ask, that is a 16-dimensional model. For each agent and each day, we take $\delta_{i,\alpha}$ in Eq. (8) as the average mid-price change determined by orders of type $\alpha$ by agent $i$. This choice is motivated by the need of keeping the overall model dimension manageable, indeed considering separate components depending on the magnitude of the price change would be impractical[1]. We choose to decompose the kernels $\phi^{i,\alpha;j,\beta}(t)$ over a set of $L$ exponential functions so that it has the form:

$$\phi^{i,\alpha;j,\beta}(t) = \sum_{\ell=1}^{L} \alpha_\ell^{i,\alpha;j,\beta} \beta_\ell e^{-\beta_\ell t}. \tag{19}$$

Furthermore, in order to limit the effects of intraday seasonality, we employ a piecewise constant baseline intensity. In particular we divide the considered trading period (8.5 hours) into 17 30-minutes intervals. The intensity of our Hawkes model has thus the form

$$\lambda_t^{i,\alpha} = \mu_t^{i,\alpha} + \sum_j \sum_{\beta \in \mathcal{T}} \int_{-\infty}^{t} \sum_{\ell=1}^{L} \alpha_\ell^{i,\alpha;j,\beta} \beta_\ell e^{-\beta_\ell(t-s)} dN_s \tag{20}$$

where the parameters of the model are the $\alpha_\ell^{i,\alpha;j,\beta}$ and the 1-day periodic piecewise baseline function $\mu_t^{i,\alpha}$, while $L$ and the $\beta_\ell$ are fixed hyper-parameters. This formulation has the advantage of being very flexible and at the same time it allows the loss function to be convex, thus greatly simplifying the estimation We estimate the model for each agent and each day separately using the contrast function method (Hansen et al. (2015), Bacry et al. (2015)) implemented in the `tick` python library (Bacry et al. (2017)). We fix $L = 10$ and we take $\beta_\ell = \frac{1}{\tau_\ell}$ with the $\tau_\ell$ taken on a log-spaced grid between $10^{-6}$s and 1s[2]. It is important to note that we do not require the parameters $\alpha_\ell^{i,\alpha;j,\beta}$ to be positive. This allows us to accommodate more flexible shapes for the kernels and to reproduce some inhibition effects that are well known to be present in financial order flows (Lu and Abergel (2017), Achab et al. (2018)).

Hence, for each day in our sample we estimate 17, 16-dimensional models, one for each of the 16 members considered plus one for all the other members combined. For each day we then construct the final $8M \times 8M$, with $M = 17$, kernel matrix by using the procedure described in Section 3. Namely, we extract all the parameters $\alpha^{i,\alpha;j,\beta}$ where the index $i$ belongs to the agent from each of the $M$ agent vs. market models. We then use these values to fill up the $8M \times 8M$ kernel matrix $\mathbf{\Phi}$ and then estimate the matrix $\mathbf{R}$ using Eq. (3). Note that these values are sufficient thanks to the

---

[1]The choice of the average is natural as

$$\Delta P = \sum_{k=1,2,\ldots,} \delta_k \Delta N^k = \Delta N \sum_{k=1,2,\ldots,} \frac{\delta_k \Delta N^k}{\Delta N} = \Delta N \bar{\delta}$$

[2]We tried several combinations of the hyperparameters and we found this one to represent the best compromise in terms of number of parameters and good description of the data. The results presented in the following sections are however quite stables over the set of hyper parameters we tried

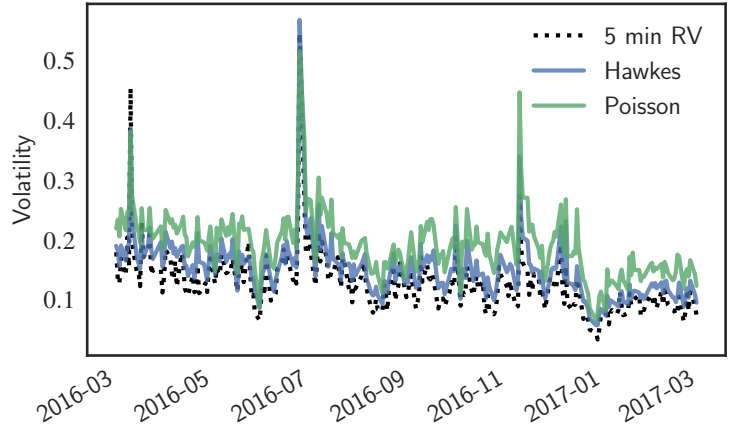

Figure 1. Comparison of 5-minute Realized Volatility estimation with volatility estimated with the Hawkes model and with a pure Poisson model.

simplifying assumption (17) we made. If one of the 16 members we consider is not present on a particular day we set all its parameters to 0, which in practice amounts to reducing the dimension of the final kernel matrix. The average baseline intensities $\mu^{i,\alpha}$ for the final $8M \times 8M$ model are then recovered using the relation (4).

To assess the goodness of our volatility decomposition, we compare (i) the daily volatility estimated with our Hawkes model using Eq. (12) with (ii) the standard daily Realized Volatility estimator based on the 5-minutes returns (Liu et al. 2015) as well as with (iii) the daily volatility estimation derived from a 2-dimensional Poisson model that uses two independent homogeneous Poisson processes for modeling respectively positive and negative mid-price jumps. In Figure 1, we plot the so-obtained estimations[1]. We note that the pure Poisson model greatly overestimates the realized volatility. According to the discussion of the toy model of Sec. 2 (see. Eq. (14)), this indicates that there are strong cross-excitation effects between upward and downward price movements. Since it can precisely account for such anti-correlations, we see that the Hawkes model performs much better albeit it also tends to overestimate historical volatility. This could be due to the hypothesis we make about the kernel matrix structure as well as to the particular choice of kernels we make, indeed a fully non-parametric Hawkes model would match exactly the volatility as it would perfectly reproduce the empirical covariance (5) (Bacry and Muzy (2016)).

We now examine how different agents and order types contribute to the total volatility. To this end, we isolate different measures from equation (12). Since the pair ($i \in \mathcal{A}, \alpha \in \mathcal{T}$) denotes one specific component of the Hawkes model, according to (12), the total volatility can be written as:

$$\sigma = \left( \sum_{i \in \mathcal{A}, \alpha \in \mathcal{T}} \Lambda^{i,\alpha} \xi_{i,\alpha}^2 \right)^{\frac{1}{2}}, \qquad (21)$$

---

[1] To convert the Hawkes squared volatility estimation $\sigma_H^2$ calculated as in (12) to the annualized value we use the formula

$$\sigma = \sqrt{\sigma_H^2 * T} \cdot \frac{S}{P_0}$$

where $S = 0.25$ is the half-tick size, $P_0$ is the open price of the day, and $T = 8.5 \cdot 3600 \cdot 252$ is the factor to annualize.

| $P^+$ | $T^a$ | $C^a$ | $L^a$ | $L^b$ | $C^b$ | $T^b$ | $P^-$ |
|---|---|---|---|---|---|---|---|
| 0.77 | 0.47 | 0.14 | 0.17 | 0.17 | 0.14 | 0.47 | 0.79 |

Table 3. Average volatility (in half-ticks) per event over the model components, averaging over all agents in Eq. (22).

where $\Lambda^{i,\alpha}$ is the occurrence rate of orders of type $\alpha$ placed by agent $i$ while

$$\xi_{i,\alpha} = \left| \sum_{j \in \mathcal{A}} \sum_{\beta \in \mathcal{T}} \delta_{j,\beta} R^{j,\beta;i,\alpha} \right| \qquad (22)$$

corresponds to the average volatility caused, in the Hawkes process sense, by one event of type $\alpha$ placed by the agent $i$ (where we assume that $\delta_{j,\beta} = 0$ if $\beta \notin \{P^+, P^-\}$).

This last quantity gives the average direct and indirect contribution of an action of type $\alpha$ by some agent $i$ to the total volatility. However, it does not fully account for the "impact" of this agent on the price fluctuations. Indeed, within the picture of Hawkes interactions, the arrival rates of orders placed by other agents $j \neq i$ also depend on the activity of agent $i$. In order to fully account for the footprint of an agent activity on the total volatility, one can compute the difference between the actual squared volatility (i.e., the squared volatility of the full model) and the squared volatility of the model without the agent $i$. We can thus define $\rho_m$ as the fraction of $\sigma^2$ that remains once the squared volatility associated with all events that are not directly or indirectly triggered by agent $m$ has been removed, all other things being equal [1]

$$\sigma^2 \rho_m = \sigma^2 - \sum_{i \neq m} \sum_{j \neq m} \sum_{\alpha,\beta} \mu^{j,\beta} R^{i,\alpha;j,\beta} \left( \sum_{j \neq m} \sum_{\beta} \delta_{j,\beta} R^{j,\beta;i,\alpha} \right)^2 . \qquad (23)$$

Let us remark that due to its non-linear dependence on $R$ it is not true that $\sum_m \rho_m = 1$. Intuitively, one can interpret this measure as the relative difference in squared volatility we would observe if we removed all the activity directly or indirectly generated (in the Hawkes process sense) by agent $m$ while leaving all the rest unchanged. Note that this quantity can also take negative values, meaning that the considered agent activity tends to stabilize the market.

### 5.1. *Volatility per Event*

In Table 3 we report the average volatility per event across all days and agents (averaging Eq. (22) on all agents). We notice that, as expected, orders that directly move the mid-price have the largest average impact on the long range volatility, followed by aggressive orders that do not move the mid-price. This result provides further evidence that our model constitutes a reasonable description of the system.

In Figure 2 we plot the average volatility per event (Eq. (22)) for each of the four event types considered and for each agent (bid and ask sides are aggregated as well as upward and downward moves, the 4 types are thus $L$ for limit orders, $C$ for cancel orders, $T$ for market orders and $P$ for price moves). We also provide values obtained from a control group of artificially created agents for comparison. Each real agent is associated to a group of 10 independent artificial agents created by randomly permuting the agents labels so that orders are randomly assigned during each day. However, care is taken to preserve the original agents' order type characteristics, so that the proportion of each order type are the same for the artificial and original agents. Each control value is then obtained by averaging on the values of each artificial agent. For all event types, we remark that there are notable differences between the true and the control values, that latter being more homogeneously distributed among agents than the former. This means that agent-specific characteristics such as the timing of orders (related e.g. to execution optimization or inventory constraints) play an important role.

---

[1] Notice that, in full rigor, the matrix $R^{i,\alpha;j,\beta}$ for $i, j \neq m$ depends on agent $m$ features and one should recompute a new $\mathbf{R}$ matrix without this agent contribution. However it is easy to show that generically, the influence of $m$ on the matrix $\mathbf{R}$ is a second order perturbation, i.e. of relative magnitude $\sim M^{-2}$.

We can explore the relationship between the volatilities per event and some agent features as chosen in section 4 (see Tables 1 and 2). We first empirically estimate, among all trading days of the database and among all agents, the feature distributions and their quantiles. This allows us to estimate the mean of $\xi$ conditionally on some agent feature being in its q-th quantile. The quantiles are computed considering each agent, each day as a separate observation. For the sake of simplicity, we did not conduct this analysis on all the 16 features of Table 1, instead we selected what we thought to be the 4 most informational features: Presence, Order lifetime, EOD Position, and Aggressive fraction. Note that in all following tables and figures we will use the complement of Presence, namely $1-$ Presence, so that increasing values in all the four features tend to mark the transition from "market maker" like to slower, directional agents. In Figure 3, these estimated conditional means are plotted as a function of the percentile index $q$ for each of the 4 features and for each of the 4 order types (cancel, market, limit and moving price). We can first remark that the volatilities per price moving events (i.e., $\xi_{i,\alpha=P}$) appear to be smaller for agent that can be qualified as high-frequency market intermediaries, i.e. fast, with mostly flat closing positions and highly present at the best limit. Some relationship with the fraction of aggressive volume appears to be present as well, with more aggressive agents presenting higher $\xi$ for price movements. Similar considerations apply for $\xi_{i,\alpha=L}$ and $\xi_{i,\alpha=C}$ associated respectively with limit and cancel orders: these are higher for slower, directional agents and lower for fast, flat agents. Finally we see that $\xi_{i,\alpha=T}$ associated with market orders does not present particularly clear patterns in relation with agents' features. This seems to indicate that the impact of a market order on volatility does not depend on the type of agent, e.g., when high-frequency market intermediaries use aggressive orders they are not "as careful as usual" and they have the same impact (per event) on volatility as an aggressive agent.

### 5.2. *Agent impact on volatility*

The quantity $\rho_m$ can be understood as the relative decrease in total squared volatility that would be observed if the agent $m$ and all the activity where he's implied were removed and the other agents did not change their activity. Clearly, the idea that other agents would not change their behavior can be rather far fetched, nevertheless, this measure is useful to provide a quantitative measurement of how much a certain agent is implied in overall volatility generation. In order to define a standardized agent influence on the market volatility, one can consider the squared volatility impact per order, i.e. $\rho_m \sigma^2 / N_m$ where $N_m$ is the total number of orders placed by the agent $m$. Along the same lines as for the volatility per event, one can estimate the average of $\rho_m \sigma^2 / N_m$ over all days and agents conditional on the agent behavior features. In Fig. 4 the conditional mean $\rho \sigma^2 / N$ is displayed as a function of the four feature percentiles considered previously. We note that all the curves show, in a strikingly consistent way, that "market-markers", i.e. agents which behavior is less aggressive and less directional, have a smaller impact per order than aggressive agents who trade less frequently in a more directional way. This is consistent with the results obtained in the previous section.

In Figure 5, we compare the squared volatility impact ratio of each agent with the control value obtained as described in the previous section. This allows us to disentangle in agent activity, the "quantity" and the "quality" of its orders. In the left plot the average value of the impact ratio $\rho_m$ for each of the agent is reported alongside the corresponding values form the control group. We observe that in many cases the actual values are either significantly higher or smaller than the corresponding control values. We also find that control values are invariably positive, while actual

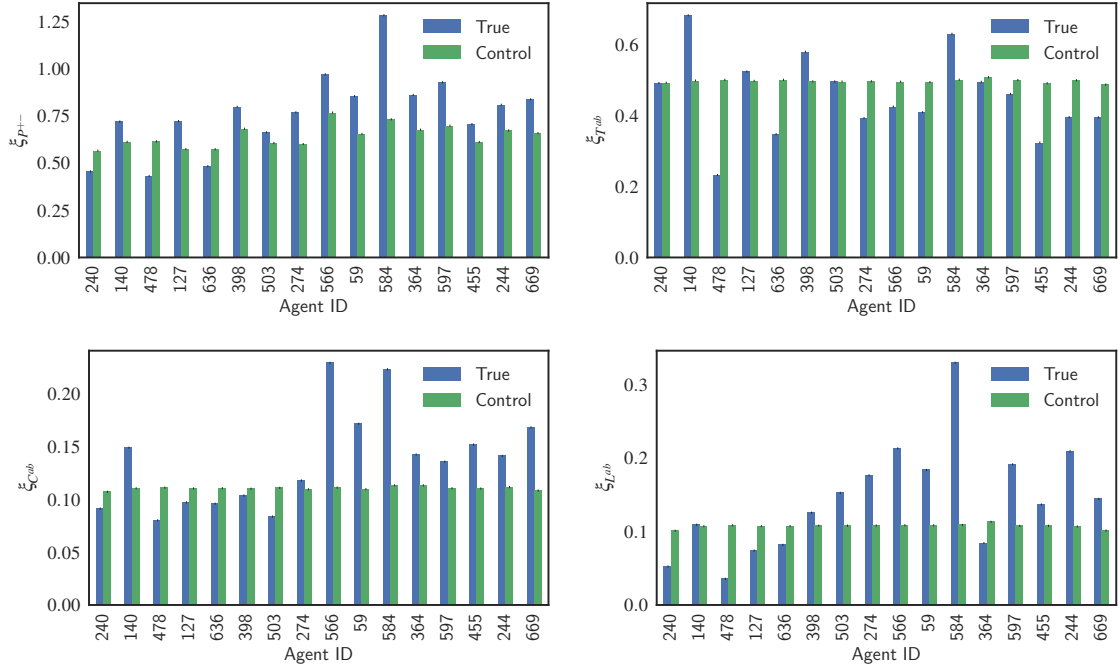

Figure 2. Average volatility (in half-ticks) per event as defined in (22) for each agent, together with the corresponding control value. The control value is an average of 10 values obtained using 10 independent artificial agents whose orders are the same as the original agents except that they are shuffled independently every single day. The four graphs correspond to the four types of events built by merging bid/ask events of the same time ($L^{ab}$ for any limit orders, $C^{ab}$ for any cancel orders and $T^{ab}$ for any market orders) and upward/downward price moves events ($P^{+-}$ events)

values attain in some cases negative values. We first notice that, as expected, $\rho_m \sigma^2$ depends on agents size, meaning that if two agents have similar profiles in terms of the response matrix $R$, then the one with more orders will have a higher impact on the total volatility. This partly explains the huge value found for the member 140, who is responsible for the largest fraction of the activity. Similarly, the total number of orders and the response matrix $R$ being equal, the order type mix also plays a role. As we have emphasized in Section 5.1, $P$ and $T$ event types have on average a greater impact on volatility. To account for these factors while comparing agents we resort once again to the control agents described before. With the artificially created control agents, we maintain the same number of events and the same event type decomposition as the true counterparts. On the other hand, the order timing and other characteristics such as the order size are randomized. In the right plot of Figure 5 we investigate the relation of the residuals $\rho_m^{\text{actual}} - \rho_m^{\text{control}}$ with some agent features such as the position at the end of the trading day and the presence at the best quotes. We note that profiles resembling high-frequency market makers present negative residuals meaning that their strategies have a lower impact on the squared volatility compared with a random strategy with the same types and number of orders.

### 5.3. *High-volatility days and Reflexivity*

Let us now investigate what happens on high volatility days. We note that, within our model, a higher value of the total volatility can be due to a change either in (i) exogenous behavior, through an increase in the baseline intensities $\mu$ or in (ii) endogenous behavior, through an adequate change in the interaction kernels and thus the matrix $\mathbf{R}$. In the following we try to differentiate between these two effects and to quantify their level. To this end, we rewrite Eq. (12) (along with Eq. (4))

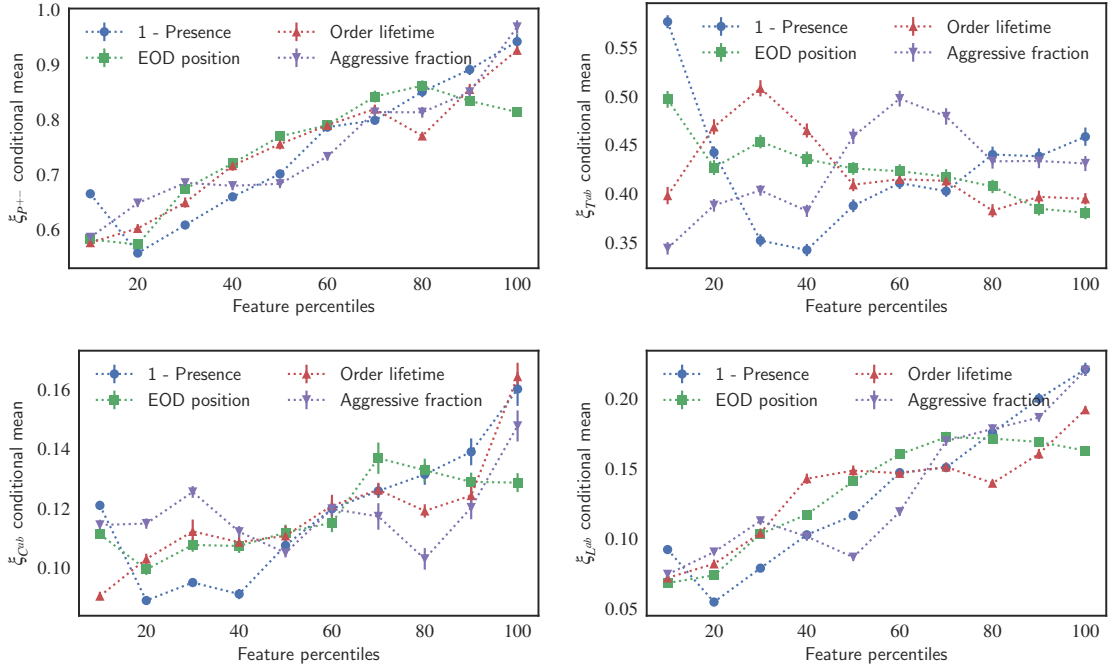

Figure 3. Conditional means of volatility per event (in half-ticks) as defined in Eq. (22) when the corresponding feature (among 4 features selected in Tables 1 and 2) is restricted to its $i$-th decile. Error bars represent $\pm$ one standard error for the mean. The four graphs correspond to the four types of events built by merging bid/ask events of the same time ($L^{ab}$ for any limit orders, $C^{ab}$ for any cancel orders and $T^{ab}$ for any market orders) and upward/downward price moves events ($P^{+-}$ events)

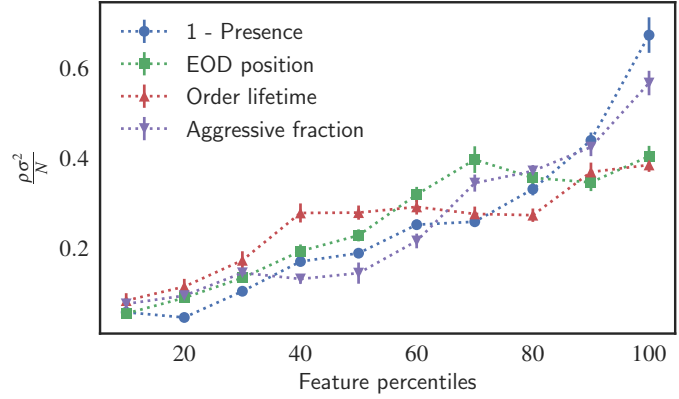

Figure 4. Conditional means of agent volatility impact per event $\rho\sigma^2/N$ when the agent feature is restricted to its $i$-th decile.

as

$$\frac{\sigma^2}{\tau} = \sum_{i,\alpha} \mu^{i,\alpha} u^{i,\alpha} \qquad (24)$$

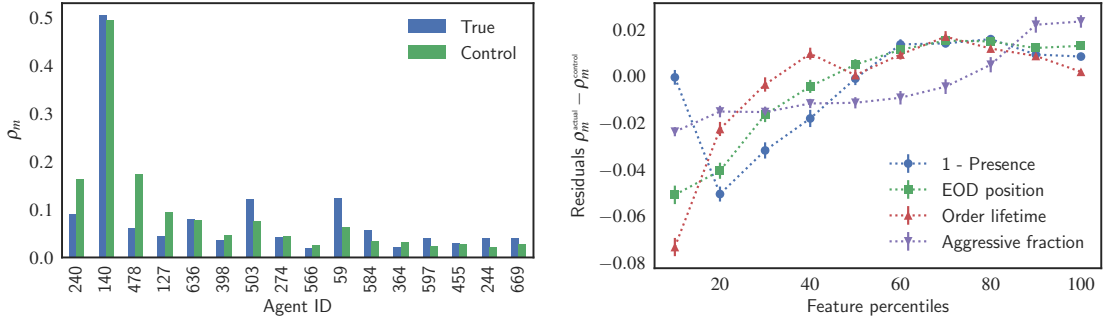

Figure 5. Left: Average $\rho_m$ for each of the considered member alongside the corresponding control value. Members are sorted from left to right according to their average EOD position, with values close to zero to the left. Right: Conditional means of the residuals $\rho_m^{\text{actual}} - \rho_m^{\text{control}}$ when the feature is restricted to its $i$-th decile. Error bars represent $\pm$ one standard error for the mean.

where $u^{i,\alpha}$ contains all the terms involving the reaction matrix $\mathbf{R}$, i.e.,

$$u^{i,\alpha} = \sum_{j,\beta} R^{j,\beta,i,\alpha} \left( \sum_{k,\gamma} \delta_{k,\gamma} R^{k,\gamma,j,\beta} \right)^2 \tag{25}$$

In order to assess how good is an approximation where the change in volatility for a given day is uniquely due to a change in the baseline intensity $\mu$ (i.e., exogenous behavior), for each day $t$ we compare

- the daily ($\tau = 1$ day) squared volatility $\sigma_t^2$ estimated from the daily estimations $\mu_t^{i,\alpha}$ and $u_t^{i,\alpha}$ using Eq. (24)

$$\frac{\sigma_t^2}{\tau} = \sum_{i,\alpha} \mu_t^{i,\alpha} u_t^{i,\alpha} \tag{26}$$

- with a version of this volatility (noted $\sigma_{\mu,t}^2$) where $u_t^{i,\alpha}$ has been replaced by an average value taken over a month (20 working days) centered at $t$, i.e.,

$$\sigma_{\mu,t}^2 = \sum_{i,\alpha} \mu_t^{i,\alpha} \bar{u}_t^{i,\alpha} \quad \text{with} \quad \bar{u}_t^{i,\alpha} = \frac{1}{20} \sum_{s \in [t-10, t+10]} u_s^{i,\alpha} \tag{27}$$

In the left plot of Figure 6, we plot the the ratio $\frac{\sigma_t^2}{\sigma_{\mu,t}^2}$ as a function of the day $t$. One can see that, with the exception of a few days that correspond to very high volatility days, the ratio fluctuates around a constant value close to one. This means that most of the volatility dynamics is captured by the variations of the exogenous rates $\mu_t^{i,\alpha}$. The same kind of conclusion was drawn in Bacry and Muzy (2014) where the authors remarked, using a non-parametric estimation of a Hawkes model for level I book events over various assets, that the U-shaped intraday volatility pattern is mainly accounted by the intraday variability of the baseline intensities while the Hawkes kernels were observed to be quite stable during the trading days.

However, it seems that this picture is not valid on very high volatility days such as the day following the Brexit referendum (24th June 2016) which corresponds to one of the large spikes in the left plot of Figure 6. This observations is quantitatively confirmed in the right plot of Figure 6 where we display the average ratio $\frac{\sigma_t^2}{\sigma_{\mu,t}^2}$ on the corresponding quantile of the daily volatility. We clearly see that the ratio is close to one, i.e., the volatility is explained by the level of exogenous intensities whatever the volatility level, except for extreme values (the 5% highest values at the extreme right of the plot). In that case, accounting for the increase of baseline intensity values is no longer sufficient to reproduce the observed volatility and it seems that these periods of large price fluctuations are characterized by a higher level of endogenous behavior.

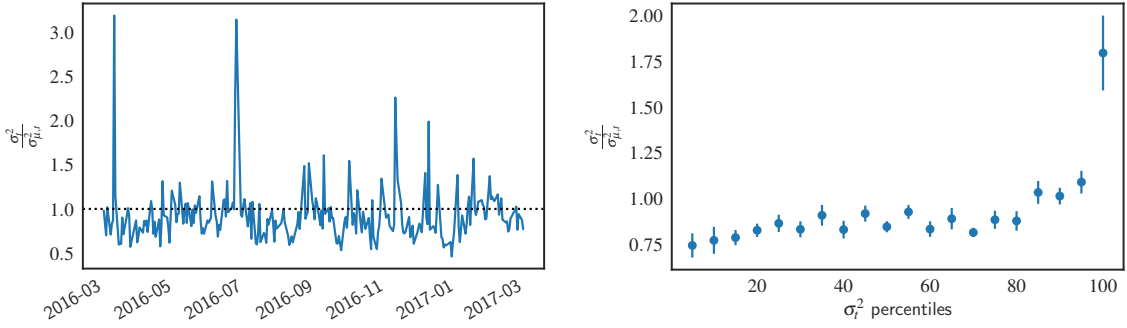

Figure 6. Left: ratio between the actual squared daily volatility Eq. (24) and the squared daily volatility as defined by Eq. (27) as a function of the day. Right: Average value of the ratio $\frac{\sigma_t^2}{\sigma_{\mu,t}^2}$ conditioned on the total volatility level.

The same kind of conclusion has been drawn in order to understand the famous "excess volatility puzzle" (Bouchaud (2018)) according to which information flow and the variations of fundamentals are not sufficient to explain the overall level of volatility. As argued notably by (Soros (2003)), a large fraction of the activity in financial market appears to be somewhat self-referential, meaning that it stems from agents reaction and feedback loops to other agents' actions rather than new information on the traded security (see also Bouchaud et al. (2018)). Hawkes processes have been used in several studies (Filimonov and Sornette (2012), Hardiman et al. (2013)) to quantify the degree of reflexivity thanks to their branching properties that allow one to separate exogenous and endogenous contributions. Thanks to the access to labeled data, we can study the difference between various types of agents as respect to their estimated fraction of exogenous activity. For that purpose, let us define the exogenous fraction $f_i$ for agent $i$ as

$$f_i = \frac{\sum_\alpha \mu^{i,\alpha}}{\sum_\alpha \Lambda^{i,\alpha}}. \qquad (28)$$

In Figure 7, we plot the averages of $f$ obtained in groups that differ by the agent features we used in previous sections like the relative absolute position at the end of the trading day, the median order lifetime or the presence at the best limit. We note that fast agents with flat end of day positions and highly present at the best limit tend to have a lower exogenous fraction. This is consistent with the idea that high-frequency traders with a target neutral position mainly exploit opportunities arising from market microstructure signals or short-lived arbitrage opportunities, thus relying almost entirely on the market activity itself to direct their actions. On the other hand, investors with longer time horizons are more likely to be driven by external information, such as proprietary economic research, and thus their activity is more "exogenous".

6. **Concluding remarks**

By leveraging the properties of Hawkes point processes and thanks to the access to labeled data, we build a model linking the high frequency behavior of market participants to their contribution to the long term volatility. Our approach can be viewed as an attempt to bridge the gap between the stochastic "zero intelligence" models and the mainstream economic approaches based on optimal decisions of fully rational, more or less informed agents (Parlour (1998), Foucault (1999), Rosu

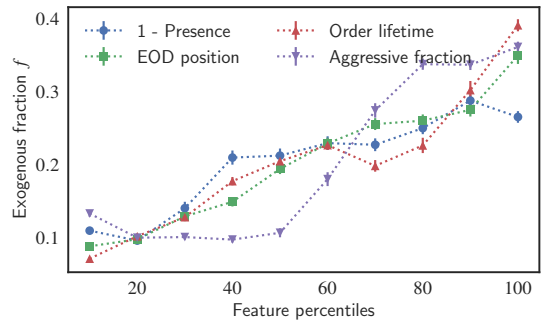

Figure 7. Conditional means of the exogenous fraction (28) when the characteristic is restricted to its $i$-th decile. Error bars represent $\pm$ one standard error for the mean.

(2009)).

We can summarize our main findings as follows. First, with regards to how different agents behaviors affect volatility, our results suggest that fast intermediaries have a stabilizing effect on the volatility, in the sense that their contribution to the overall volatility is smaller than what could be implied by the scale of their activity. This is not the case for more aggressive, directional agents that have a significantly larger impact per action on the price volatility. Second, we investigated, within the framework provided by our model, what factors determine changes in volatility on a daily time scale. We observe that variation in the level of volatility can be chiefly ascribed to variation in the "baseline" activity of market participants, rather than in changes to the way market participants react to the order flow. Interestingly, this does not appear to hold on days of severe market turbulence where we find that also the reaction component changes markedly. Moreover, by comparing the level of endogeneity between different typologies of market participants we confirm the results of several studies (see Bouchaud (2018) and references therein) that a majority of market activity can be explained by self-reflective mechanisms, and in particular our study reveals that this is particularly true for market participants typically employing order-book based algorithmic strategies as it is the case for high-frequency market makers.

Finally, let us mention that, as the Hawkes kernel matrix encodes a great deal of information on the agent, it could be used as input to clustering algorithms in order for example to group agent with similar behaviors or to detect changes in the behavior of a particular agent. This is similar in spirit to Yang et al. (2015), where the authors use inverse reinforcement learning to infer the reward functions of different agents and then use these as input features for a clustering algorithm. Our approach lacks the strategic view of agent behavior associated with the value optimization framework, however it also requires significantly less data in order to produce an estimation allowing to obtain an independent estimate for each day and includes information on order timing by working in physical (continuous) time rather than event time.

## Acknowledgments

We thank Euronext for making their data available to us. We also thank Laurent Fournier, Angelique Begrand and Luxi Chen for useful discussions. This research benefited from the support of the chair of the Risk Foundation: Quantitative Management Initiative.